\documentclass[prb,aps,superscriptaddress,endfloats, showpacs]{revtex4}
\usepackage{epsfig}

\begin {document}

\title {About low field memory and negative magnetization in semiconductors
and polymers.}

\author {I. Bulyzhenkov} 
 \altaffiliation[Also at ]{the Institute of Spectroscopy RAS, Troitsk, Moscow reg. 142092, Russia}
\email {ibuly263@science.uottawa.ca}
\author { A.-M. Lamarche} 
\author {G. Lamarche}
\email {glamarch@science.uottawa.ca}

\affiliation{Ottawa-Carleton Institute of Physics, University of Ottawa,  \\ 150,  Louis-Pasteur, Ottawa, Ontario K1N 6N5, Canada}

\begin {abstract} 
 Ginzburg-Landau bulk magnetization of itinerant electrons can provide a negative effective field in the Weiss model by coupling to localized magnetic moments.  The coupling enforces remnant magnetization, which can be negative or positive depending on the sample magnetic history. Stable magnetic susceptibility of coupled nonequilibrium subsystems with magnetization reversal is always positive. Gauss-scale fields could be expected for switching between negative and positive remnant moments in semiconductors with coupling at ambient temperatures. Negative magnetization in ultra-high conducting polymers is  also discussed within the developed framework. 
\end {abstract} 

 {}

\maketitle

\section {Introduction }

Observations at low temperatures of negative magnetization in semiconductors with transition metal atoms have been reported \cite {Men,Sak} since 1960, when asymmetric hysteresis loops at 4.2K were found in cobalt vanadate and cobalt titanate. Antiferromagnetic perovskites with rare earths, for example \cite {Good,Yosh, Yelon} with La, Pr, and Ho, also exhibited "anomalous diamagnetism" or reversal of magnetization. The thermal behavior of negative magnetization in many inorganic and organic structures is still under discussion on the qualitative level. Most authors tend to relate this unusual phenomenon exclusively to one\cite {Men,Good} or two\cite{Shin} compensation points, envisaged by N{\'e}el  for ferrimagnetic systems\cite {Neel}.   

However, all compensation points are always below the long-range ordering, while "anomalous diamagnetism" has also been measured \cite {LLL} well above the N{\'e}el temperature. A higher melting rate of stronger spin sublattices in antiferromagnetic structures could be assumed for transitions to negative remnant moments under warming of semiconducing selenides\cite {Sny}, for example. But could weaker sublattices  originate primarily with negative moments under reverse cooling either in monocrystalline or in polycrystalline samples, including vibrating powders in VSM magnetometers? Moreover, magnetization reversal in amorphous  organic samples \cite {Eni}  would unlikely be explained through N{\'e}el's compensation for ordered subsystems.

 Many verified observations gave liberty to assume even diamagnetic grounds  for negative magnetization in semiconductors \cite {Shira,Cla}and ultra-high conducting polymers \cite {Eni,Rog} because it was difficult to interpret this unusual phenomenon through responses with positive magnetic susceptibilities. Nonetheless, we intend to suggest that moment reversal 
 in any substance does not manifest grounds for diamagnetism unless negative slopes of closed hysteresis loops are registered.    

The goal of this paper is to propose a phenomenological approach to negative bulk magnetization when it is maintained by coupling to localized moments of paramagnetic ions or polaron-type magnetic aggregates. The initial microscopic origin of the weak "molecular field" moment is not of much importance for a phenomenological model. The bulk interatomic moment may be related, for example, to delocalized covalent states in molecular formula units or to other kinds of itinerant electrons. We just intend to prove that magnetic coupling to disordered or ordered localized moments can universally enforce and maintain negative bulk magnetization in the zero-field limit. At the same time all coupled subsystems may contribute to the total susceptibility with only paramagnetic-type responses on applied fields.     

When a delocalized valence electron is engaged in interatomic bonds, for example, then its magnetic moment ${{\bf m}} = - (\mu_{_B}/ \hbar ) ({\bf l} + {\bf s}) - 
 m(\mu_{_B}/ \hbar )^2(x^2 + y^2) {\bf h}$ depends on angular momentum ${\bf l}$ and spin $\bf s$. Hereinafter $\mu_{_B} = e\hbar / 2m$ is the Bohr magneton. The last item in ${{\bf m}}$ is always antiparallel to the external magnetic field applied along the z-axis, ${\bf h} = h_z = h$. This term corresponds to diamagnetism and negative contribution to the total  susceptibility. However, diamagnetic induction vanishes in zero fields and cannot be responsible for negative remnant moments.  

Dynamical correlation of the delocalized valence electron with two or more covalent partners from neighboring equidistant atoms  may improve,  for example, charge distribution symmetry in the molecular unit. Spontaneous transition at some critical temperature to higher symmetry with dynamical or rotating bonds would release potential energy for interatomic charge density circulation that could result in bulk magnetization even in dielectrics.  We just postulate that delocalized or itinerant electrons can create weak magnetization within $N_f$ formula units of a macroscopic sample with $N = K N_f$ magnetic ions. And we assume that the average bulk moment,  $ {\bf M}(T,h)$ = ${N<{\bf m}>}$, of delocalized electrons may keep finite values even in the zero-field limit, when $h \rightarrow 0$. Then we derive how magnetic coupling with ions can maintain and enforce  this bulk  moment in paramagnetic and antiferromagnetic subsystems.   

We also discuss options to switch negative remnant moments into positive ones within the framework of the Ginzburg-Landau (GL) model, which admits magnetic memory mechanism even for very low fields. Experimental opportunities to test the developed approach to negative remnant magnetization in inorganic and organic conductors with magnetic coupling will be considered.

\section {Bulk magnetization coupled to paramagnetic moments}

Any magnetic ion with total angular moment of electron shells has in general a localized magnetic moment $m_J = (- g)\mu_{_B} J$ (hereinafter $\hbar$ = 1 for short). The Land\'{e} factor $(-g)$ with a negative sign for electrons, when $g = [3J(J+1)+S(S+1)-L(L+1)]/2J(J+1)$, involves the spin, S, and the orbital, L, numbers for the net angular moment J of localized states.  Crystal fields in solids lift the d-shell degeneracy and quench, as a rule, the orbital moment L. Therefore one may relay the localized magnetic moment to the spin $S$ of the 3d shell and use $m_S = -2\mu_{_B}S$ for sites with transition metal ions, by keeping  general Land\'{e} relations for rare earth or diamagnetic ions, if any.

Would interatomic magnetization spontaneously arise for some reason in molecular formula units or polaron-type aggregations of charged counter-partners, then localized magnetic moments start to interact or couple to this magnetization. Again, details of microscopic origin for bulk interatomic magnetization is not a matter for a phenomenological approach.  Important is that the bulk z-axis moment, $M(T,h)= M$, of any origin is to be coupled to localized magnetic moments in agreement with their contribution to the system  Hamiltonian, 

\begin{equation}
	{\cal H}'=N_f\sum_{n=1}^K\left (g_n\mu_{_B} J_n h + \Lambda_n J_n {M\over N \mu_{_B} }\right )- M h,               
\end{equation}
where $\Lambda_n = \Lambda_n (T) $ is the phenomenological parameter for coupling of the angular moment $J_n$ (or spin $S_n$ for 3d shells) of the "n"-th formula's ion
and the dimensionless bulk moment, $M/N \mu_{_B}$, averaged per one magnetic ion. Recall that every formula unit contains K ions with different or similar magnetic moments. We may also accept the formula unit notion for organic substances with magnetic polarons or other low mobile aggregates with magnetic moments.

One can derive from (1) the effective magnetic field,
\begin{equation}
	H_n(h,M)= h + {{\Lambda_n M (T,h)} \over g_n N \mu^2_{_B}}, 
\end{equation}
 which controls behavior of the localized moment $J_n$ in the "n"-th formula unit site.  It is essential that this z-axis field can be negative, $H_n < 0$,  in z-positive applied fields, $h >0$, when $M(T,h) < 0$ and $\Lambda_n > 0$. Such universal opportunity for the ion subsystem can result in the nonequilibrium reversal of their paramagnetic moments in low magnetic fields below some transition temperature, which admits coupling of magnetic subsystems.

The value of the subsystem magnetization $M/N$  
can be found, in principle, from the dynamical equilibrium condition, $\delta G/ \delta M = 0$, for the total system of magnetic moments.
The  Gibbs free energy  $G$ of the macroscopic sample depends   on the Helmholtz magnetic potential  $F(M, T)$, the bulk moment energy in the applied magnetic field, and the ensemble energy of localized ion moments in the effective field (2), respectively,  
\begin{eqnarray}
	G= G_o+ F(M, T)-Mh
	 - N_fk_{_B}T \sum_{n=1}^K   ln {{sinh[(2J_n + 1)p_n/2J_n]}\over sinh (p_n/2J_n)},
\end{eqnarray}
where $p_n \equiv g_n\mu_{_B}J_nH_n(h,M)/k_{_B}T$ is the dimensionless parameter of the Weiss molecular field model \cite {Stan,Blo}.  

Now we employ the general Landau presentation,  $F(M, T) - F(0,T) =  2^{-1}a(T) M^2 + 4^{-1}b(T) M^4 + 6^{-1}c(T)M^6 + ... $, for magnetic shifts of the Helm\-holtz free energy. Then the dynamical equilibrium condition for the bulk moment $M$ reads as follows
\begin{eqnarray}
a(T) M  + b(T)M^3 + c(T)M^5 + ... 
 = h + N_f\sum_{n=1}^K{{\Lambda_n J_n}\over N\mu_{_B}} { B} (J_n, p_n),
\end{eqnarray}
where ${ B}(J,p) = [(2J+1)/2]coth [(2J + 1)p/2J] - (2J)^{-1}coth (p/2J)$ is the Brillouin function. Recall that ${ B}(J_n,p_n)\rightarrow 1 $ for $p_n \gg 1$ (T $\rightarrow$ 0)
 and ${ B}(J_n,p_n) \approx p_n(J_n+1)/3J_n = g_n(J_n+1)\mu_{_B}H_n/3 k_{_B}T$  for $p_n \ll 1$ ($ H_n \rightarrow 0 $). The latter limit ordinarily takes place in the laboratory, because $\mu_{_B}H_n \ll k_{_B}T$  can be accepted up to tesla-scale fields for temperatures above 4.2K. 
The right hand side of (4) is to be considered as an effective magnetic field, which 
controls the bulk moment $M$ and, consequently, the averaged magnetization $M/N$. This effective field, like the field (2) for localized moments, is shifted with respect to the applied field $h$ due to mutual magnetic coupling of localized and delocalized (valence or itinerant) electrons. In other words the equilibrium state of the total electron system with coupling is accompanied by non-equilibrium thermodynamical states of subsystems.

Different options for solutions $M(T,h)$ may be considered in (4), depending on the Landau parameters $a(T), b(T), c(T), etc.$ Would the system obey $a(T) > 0 $, $b(T) < 0$, it might gain negative magnetization in applied field, but not equilibrium residual magnetization. There is sense to keep the third term with
positive parameter $c(T)$ at the left hand side of (4) only for special cases with $b(T) < 0$. Otherwise the system achieves equilibrium states under negligible contributions from the third and following small terms. Systems with $a(T) < 0 $ and $b(T) > 0$ originally correspond to  long-range ordering with nonzero ferromagnetic moments in  zero fields that is out of our immediate interests.       
         
Below we consider paramagnetic systems with  $a(T) > 0 $ and $b(T) > 0$, {\it i.e.}  with  a positive magnetic shift of the Helmholtz free energy. At first glance average bulk magnetization costs energy and cannot exist. However one can rewrite (4) in the equivalent Ginzburg-Landau form \cite {Lan},
\begin{eqnarray}
	   b(T){{M(T,h) } }\left [{{M^2(T,h) } } - {{M^2_o (T) } }	\right ] = ({1+ {P_{\Lambda}}}) h,
\end{eqnarray}
and to conclude that the magnetic coupling between delocalized  and  localized  electron subsystems can initiate spontaneous moments $M(T,0) = \pm M_o(T)$ even in pure paramagnetic systems.    
Here  $P_\Lambda = P_\Lambda (\Lambda_n, J_n,T)$ is the dimensionless enforcement parameter,
\begin{equation}
P_\Lambda =  {1\over K}\sum_{n=1}^K {{g_nJ_n(J_n+1)\Lambda_n}\over 3k_{_B}T},
\end{equation}
and $M_o(T)= M_o(T,\Lambda_n,J_n,a,b)$ is the absolute value of the GL spontaneous moment of delocalized electrons,
\begin{eqnarray}
 M_o(T) \equiv\left (- {{a (T)}\over b(T)} + {{1}\over K} \sum_{n=1}^K {{J_n(J_n+1)\Lambda_n^2}  \over 
3k_{_B}T N \mu_{_B}^2  b (T)   }  \right )^{1\over 2}.
\end{eqnarray}
Low field diamagnetic solutions of (5), $M(T,h) \approx - (1+P_\Lambda)h/M_o^2(T) b(T)$, are not stable in the GL model. The moment $M(T,h)$ drops from these unstable negative solutions to  stable negative solutions in the range $- M_o(T) \leq M(T,h) \leq -M_o(T)/{\sqrt 3}$ below the coupling transition temperature  
\begin{equation}
  T_\Lambda \equiv {1\over K}\sum_{n=1}^K{{J_n(J_n+1)\Lambda_n^2 }  \over 
3k_{_B}a (T_\Lambda)N \mu_{_B}^2}. 
\end{equation}
This temperature can reach 300K and more under the high coupling energies $\Lambda_n $ and the high angular moments $J_n$ of ions with partially filled 3d or 4f magnetic shells, if the positive structural parameter $a(T)$ is sufficiently small.

Cubic GL dependence of the moment $M(T,h)$ on applied fields in (5) suggests hysteresis phenomena and magnetic memory states related to interplay of two stable and one unstable equation solutions. In strictly zero fields there is no preferred direction in isotropic paramagnetic systems and the averaged macroscopic moment is absent. This distinguishes coupled paramagnetic systems from ferromagnetic structures which gain remnant moments along
 the easy axis of magnetization. The diamagnetic branch of GL solutions with small negative moments, $M(T,h) \propto - h$, is unstable even in very low fields in which the bulk moment spontaneously gains finite negative values $M(T,h\rightarrow 0) =  -M_o(T)$. Further increase of the applied field $h$ in (5) decreases the absolute value of
  the negative moment up to a threshold $-M_o(T)/{\sqrt 3}$
  at the "switching" field   
 \begin{equation}
h_s = {2{\sqrt 3  a^{3/2}(T)  } \over 9(1+ P_\Lambda) b^{1/2}(T)}  \left ({{T_\Lambda -T}\over T}\right )^{3/2}.
\end{equation}
The field $h_s$ switches the negative moment into another branch of solutions with  $ x =M(T, h\geq h_s)/ M_o\geq 1.155 $ in agreement with the local extremum $2{\sqrt 3}/9$ for
the function $x(x^2 -1)$ at the left hand side of (5).

If positive applied fields below $h_s$ are turned off, the negative bulk moment $M(T,0)$ relaxes to the spontaneous value, $-M_o(T)$. If fields above $h_s$ are turned off, the positive bulk moment $M(T,0)$ relaxes to another stable value, $+M_o(T)$. Both stable remnant moments correspond in isotropic systems to the same Helmholtz potential, $F(+M_o, T) = F(-M_o,T)$. Would the substance with $h_s >$ 0.5 Oe be warmed above the critical temperature (8) and cooled in zero or Earth's magnetic field (0.5 Oe), the stable GL moment gains again the negative spontaneous value around $-M_o(T)$. However unstable diamagnetic originations with small negative moment may take place, in principle, just after the cooling unless the system falls to the strongest negative magnetization with further paramagnetic responses to any applied fields. 

The GL mechanism of magnetic memory in isotropic systems, which depends only on values of applied fields, is different from the ferromagnetic memory mechanism, which depends on directions of fields regarding the easy axis of magnetization. 
Negative or positive magnetization of prospective materials with coupling could be recorded just by weak, gauss-scale pulses. 

The bulk moment $M(T,h)$ with two stable residual values $\mp M_o(T)$ belongs only to the delocalized electron subsystem. A total magnetic moment of the sample,   ${\cal M}(T,h) = - dG/dh$, can be derived from the Gibbs free energy (3) for the total system of coupled delocalized  and localized moments, 
 \begin{eqnarray}
 { \cal M}(T,h)=M(T,h)+{{N_f}} \sum_{n=1}^K g_n\mu_{_B}J_n { B}(J_n, p_n)         \nonumber \\ 
      \equiv (1+ P_\Lambda)M(T,h)+ {{ \sum_{n=1}^K N_fg^2_n\mu^2_{_B}J_n(J_n+1)h}\over  3k_{_B}T}.
            \end{eqnarray}
The last term at the right hand side of (10) is the regular paramagnetic contribution. This term vanishes in zero applied fields, but ion moments still maintain the zero-field moment of delocalized electrons, 
$M(T,0) = \mp M_o(T)$, through magnetic coupling with the enforcement factor $P_\Lambda$
defined by (6).

Now one can determine from (10) a magnetic field $h_0$ in which the sample moment  vanishes,    
\begin {equation}
h_0\approx{{3k_{_B}T(1 + P_\Lambda)M_o(T)}\over N_f\mu^2_{_B}\sum_{n=1}^K g^2_nJ_n(J_n+1)}.
\end {equation}
Here we used the sample remnant moment, ${\cal M}(T,0) = - {(1+P_\Lambda)M_o(T)}$, for the low field approximation, $M(T, h_0) \approx - M_o(T)$, in the non-linear equation 
${\cal M}(T, h_0)$ = 0.  

The total  susceptibility, $\chi(T,h) \equiv d{\cal M}(T,h)/ dh$, of the sample with coupled magnetic subsystems depends on the applied magnetic field,  
 \begin{eqnarray}
\chi(T,h)={{(1+P_\Lambda)^2 }\over {b(3M^2 - M^2_o)} }
 +  {{ {\sum_{n=1}^K}N_fg_n^2\mu^2_{_B}J_n(J_n+1)}\over 3k_{_B}T }
 \nonumber \\
\equiv {{(1+P_\Lambda)^2 }\over {2bM^2_o+ 3(1+P_\Lambda)M^{-1}h}}+{{{\sum_{n=1}^K} N_fg_n^2\mu^2_{_B}J_n(J_n+1)}\over 3k_{_B}T },
 \end{eqnarray}
where we used (5) and $b(3M^2 - M^2_o)dM/dh = 1 + P_\Lambda$ for $T \leq T_\Lambda$. Recall that the sign of the bulk moment $M(T,h)$ depends on the magnetic history of the sample. The magnetic susceptibility (12) for the non-linear moment (10) is positive on both stable GL branches where $M^2(T,h)\geq M^2_o(T)/3$. Metastable diamagnetic states,  $- M_o(T)/{\sqrt 3}< M(T,h) <0$, might be expected only in very low fields, if  samples have no magnetic history after zero field cooling (ZFC) below $T_{\Lambda}$. Peculiarities of the susceptibility at $M^2(T,h_s)= M^2_o(T)/3$ might be smoothed in practice due to nonhomogeneity of averaged  magnetization $M(T,h)/N$ over the sample with different magnetic ions and with different covalent radii of like ions in polyvalent states.

\section {Antiferromagnetic systems and magnetic multicoupling}

The phenomenological model for coupling of the averaged bulk magnetization $M/N$ to localized paramagnetic moments may be modified for multicoupling to various short-range and long-range ordered moments. At some temperatures the coupling might be related to ordered and disordered magnetic sublattices, while at higher temperatures only disordered moments can  enforce remnant bulk magnetization. Coupling of molecular or cluster orbital moments to the lattice of localized spins of transition metal ions might be appropriate for glass states in antiferromagnetic semiconductors \cite {Be,Fi,Go}. Another direction for the GL framework development are amorphous polymers\cite {Eni,Rog} with itinerant electrons and anomalous negative magnetization.   
   
Below we  consider how to apply the general scheme with coupling of localized and delocalized moments to antiferromagnetic systems. In this case the effective magnetic field (2) in the "{\it n}"-th ion site,
\begin{equation}
	H_n = h + {{\Lambda_n M + \lambda_{n1} m_1 + \lambda_{n2}(-m_2)  } \over g_n N \mu^2_{_B}}, 
\end{equation}
depends on the bulk moment $M=M(T,h)$ of delocalized electrons and two opposite lattice moments, $m_1=m_1(T,h)$ and $(-m_2)= -m_2(T,h)$, and relevant coupling parameters $\lambda_{n1}$ and $\lambda_{n2}$. Now the Gibbs free energy (3) reads with three bulk moments, $M, m_1,$ and $(-m_2)$, which are mutually coupled in the Helmholtz magnetic potential:   
\begin {eqnarray}
F(M,m_1, m_2,T)=F(0,T)+{{a(T)}\over 2}M^2+{{b(T)}\over 4}M^4
+{{\alpha_1(T)}\over 2} m_1^2 + {{\beta_1(T)}\over 4} m_1^4
\nonumber \\+{{\alpha_2(T)}\over 2} m_2^2 + {{\beta_2(T)}\over 4} m_2^4
 - \gamma (T) m_1m_2 + \theta_1 (T) M m_1- \theta_2 (T) M m_2. 
\end {eqnarray}

The temperature functions  $\alpha_{1/2}(T)$ and $\beta_{1/2}(T)$ determine N{\'e}el's compensation points and ferrimagnetic properties of samples. However we consider below only strictly antiferromagnetic options, for simplicity, when  $\alpha_{1/2}(T) = \alpha $, $\beta_{1/2}(T) = \beta$,   $\theta_{1/2} (T) = \theta$, and $\lambda_{n1/2} =\lambda_n$. Then
variations of the Gibbs energy (3) with (14) and (13) with respect to $M, m_1,$ and $m_2$ lead to the simplified system of three equations, respectively,

\begin {eqnarray}
{\cases {(a-\Delta)M +bM^3 = (1+P_\Lambda)h + (\delta - \theta)(m_1-m_2) \cr 
(\alpha - \delta)m_1 +\beta m_1^3  = (1+P_\lambda)h  + (\Delta-\theta)M + (\gamma -\delta) m_2  \cr 
(\alpha - \delta)m_2 +\beta m_2^3 = -(1+P_\lambda)h  - (\Delta-\theta)M  + (\gamma -\delta) m_1 \cr 
}},
\end {eqnarray}
where we defined new parameters  
$P_\lambda$=$\sum_{n=1}^K {{g_nJ_n(J_n+1)\lambda_n}  / 
3Kk_{_B}T }$,
$\Delta = \sum_{n=1}^K {{J_n(J_n+1)\Lambda_n^2}  / 
3Kk_{_B}T N \mu_{_B}^2 }$,
$\delta $ = $\sum_{n=1}^K {{J_n(J_n+1)\lambda_n^2}  / 
3Kk_{_B}T N \mu_{_B}^2 }$.

Solutions for strictly antiferromagnetic moments in (15), $(m_1-m_2) \approx [M (\Delta - \theta) +h(1+P_\lambda)]/(2\gamma - \alpha - \delta)$  and $m_1^2$ $ \approx m_2^2$ $ \approx (\gamma-\alpha)/ \beta > 0$, confirm that the remnant ferrimagnetic imbalance  $(m_1-m_2)$ can vanish only in the absence of the bulk moment of delocalized electrons, when $M(T, h)\rightarrow 0$ and $h\rightarrow 0$. Now the bulk moment can be found from the first equation (15), 
\begin {eqnarray}
 b(T)M(T,h)[M^2(T,h) - M^2_{03}(T)] =
\left [1 + P_\Lambda + {{(1+P_\lambda)(\delta -\theta) }\over (2\gamma - \alpha - \delta)}
 \right ]{{h}},  
\end {eqnarray}
where $M_{03}(T)\equiv b^{-1/2}[-a +\Delta + (\Delta - \theta)(\delta -\theta) (2\gamma - \alpha - \delta)^{-1}]^{1/2} $ is the modified amplitude for the spontaneous GL moment of delocalized electrons in coupled systems with antiferromagnetic distribution of localized moments. 
The coupling transition temperature $T=T_{\Lambda\lambda}$ is defined for such systems by an equation $M_{03}(T)=0$. The sample magnetic moment, ${ \cal M}(T, h)= -dG/dh$, contains contributions from $M$, $m_1$, and $m_2$,
\begin {eqnarray}
{ \cal M}(T, h) = \left (1+P_\Lambda +  {{(1+P_\lambda)
(\Delta -\theta) }\over (2\gamma - \alpha - \delta)}  \right)M(T,h)  
+ \sum_{n=1}^K {{N_fg^2_n\mu^2_{_B}J_n(J_n+1)}\over  3k_{_B}T}h
+ {{(1+P_\lambda)^2 
(\delta -\theta) }\over (2\gamma - \alpha - \delta)}h,
\end {eqnarray}
where we used (13) and $(m_1-m_2)\propto M$ from (15). The bulk component $M=M(T,h)$ obeys (16) and depends on the sample magnetic history. 

The total remnant moment, ${\cal M}(T, 0)$, can be negative in ZFC samples without a magnetic history when $M(T,0)=-M_{03}(T)$.
 Once at low temperatures molecular units gain their spontaneous moment, then it is enforced and maintained by both paramagnetic and antiferromagnetic moments in the macroscopic system. The temperature dependence of the last term in (17) is rather complicated and does not obey the regular Curie-Weiss law. At low temperatures the coefficient 
$(1+P_\lambda)^2(\delta -\theta)$, for example, can be proportional to $T^{-3}$, rather than to $T^{-1}$. 

By following the proposed scheme, one can derive from (3) and (14), with $\alpha_1 \neq \alpha_2$ and $\beta_1\neq \beta_2$, an analog of (15) for ferrimagnetic systems with compensation points $T_{com}$ below the N{\'e}el transition $T_N$. Five combinations for three main temperatures $T_N$, $T_{com}$, and $T_{\Lambda\lambda}$ (with its thermal hysteresis shift $\Delta T_{\Lambda\lambda} > 0$) could be considered for ferrimagnetic structures with sole compensation point and bulk molecular magnetization.
One can also consider a ferromagnetic analog of (15)-(17) for interplay of one lattice moment $m_1$ with the bulk moment $M$ by taking $m_2 \equiv 0$ in (14) and (3).  
In summary compounds with magnetic coupling and negative remnant moments are strongly non-linear systems in weak magnetic fields, where they do not comply to ordinary laws for uncoupled systems above and below the N{\'e}el and Curie transitions.

\section {Options for laboratory tests}

In general there is as much physical phenomena at room temperatures in applications of GL type equations, like (5), (15), or their extensions, to semi-, ultra-, and normal conductors with magnetic coupling of electrons as in applications of the original GL equation to superconductors with paired  electrons at low temperatures. The Earth's field limit 0.5 Oe is especially unique for applications of semiconductors with negative spontaneous magnetization and magnetic memory states. In our view magnetic coupling between localized and delocalized or itinerant electrons may reveal in practice many new and unusual phenomena, including nonequilibrium and nonstationary states with very slow relaxation rates.
Stable magnetic susceptibility is always positive in the present framework for coupled electron states. However, one may expect in very low magnetic fields that unstable diamagnetic responses can precede spontaneous formation of negative magnetization  in isotropic samples. It is a matter of experimental tests to confine encountered model options for compounds with coupling above or below points of long-range magnetic ordering, if any.

The developed approach to negative remnant magnetization in condensed matter might be tested at first with the well-studied paramagnetic or antiferromagnetic systems. For example,   it was underlined \cite {Mor} many years ago that the magnetic properties of antiferromagnetic compounds $\rm MCr_2Se_4$ (M=Fe, Co, Ni) at 200K - 270K are very unusual and do not show Curie-Weiss behavior above the N$\rm \acute {e}$el point. Therefore we assume that defect-NiAs semiconductors with transition metal atoms, like $\rm M_xCr_{3-x}A_4$ (M=Fe,Ti,  Cu, Zn, Cd, La, Pr, Sm  and A=Te,Se), and doping (Ga,Ge,As) could be appropriate candidates for comparative tests under the general approach to negative magnetization in systems with coupling. There are particular reasons to choose defect-NiAs structures with high-spin Cr cations for initial tests. First, homogeneous chalcogenides can be reliably created in the regular lab environment and many of their properties have been widely studied. Second, one can adjust\cite {Koj,Sny} a long-range ordering point from 80K in the parent compound $\rm Cr_3Se_4$ to the 350K in $\rm Fe_{2.5}Cr_{0.5}Se_4$ depending on specific investigation goals. And third, the  $\rm FeCr_2Se_4$ sample has already exhibited \cite {LLL} negative remnant magnetization up to 270K, while its N$\rm \acute {e}$el temperature was only about 190 K according to magnetic Bragg peaks in the neutron diffraction studies. Tellurides are known as slightly ferromagnetic compounds at room temperatures, therefore they might exhibit stronger coupling and magnetization reversal at helium temperatures than selenides.

Based on (10) and (17) for temperatures below $T_\Lambda$ and $T_{\Lambda\lambda}$, respectively, we can predict that even negative ZFC moments ${\cal M}(T,0)$ of samples should have positive, paramagnetic-type slopes in magnetic hysteresis curves. These slopes may be different in different field intervals according to (12), where the bulk component 
$M = M(T,h)$ can be negative or positive depending on the sample magnetic history.
Once negative or positive magnetic moment $M(T, h)$ of the delocalized electron subsystem is formed under ZFC or field cooling to $T_\Lambda$ and below, then this bulk moment is maintained  by the other subsystem with localized moments. Now the maintained or enforced moment can be destroyed by warming only above $T_\Lambda$. Therefore one has to declare a positive thermal hysteresis ($\Delta T_\Lambda > 0$ under warming) in the developed model for subsystems with magnetic coupling. This specific feature would help to distinguish at low temperatures the coupling mechanism from N{\'e}el's algebraic compensation of moments  without thermal hysteresis.

Polycrystalline samples can maintain a mixture of granules with negative and positive moments in low applied fields $h$. When a paramagnetic sample with a negative remnant moment is homogeneous and isotropic, with the symmetry  $F(-M, T) = F(+M,T)$ for potential energy, then the sample should exhibit the GL magnetic memory and hysteresis in low fields above the switching value (9).  The switching value $h_s$ can be adjusted (by doping, for example) just above the Earth's 0.5 Oe. And at room temperatures the GL interbranch switching might be expected for some magnetic semiconductors at gauss-scale fields.

Recall that the negative remnant moment, ${\cal M} (T,h \rightarrow 0) < 0$, is ultimately maintained by all coupled electrons. Slight hysteresis of pure paramagnetic systems with coupling at high temperatures and tiny remnant magnetization, $\mp(1+P_\Lambda)M_o(T)/N \approx \mp 10^{-5}\mu_{_B}$, due to the isotropic GL memory mechanism, can simulate ferromagnetic ordering of paramagnetic ions well above Curie temperatures. However, when a residual bulk moment, $\pm M_0$, is really shifted by ferromagnetic inclusions, with  $F(-M_0,M_{in},T) \neq F(+M_0,M_{in},T)$, or when there are other anisotropic corrections to the macroscopic Helmholtz potential, then the positive sample moment can be metastable with a measurable relaxation time toward the stable negative moment. Nonstationary GL generalization can be developed and applied to such metastable systems with macroscopic anisotropy. In particular, they could exhibit nonequilibrium hysteresis loops only around moderate applied fields, where GL bulk component reaches its saturation, while hysteresis shifts around $h=0$ and at high fields
would vanish. 

There are also many other relaxation phenomena which can obey the known nonstationary generalization of the GL equation, for example \cite {Tin,Schmid}. The electron-phonon relaxation  rate $\tau^{-1}\approx 10^8 s^{-1}$ for paired superelectrons is
about six orders lower than the plasma rate for free electrons in normal metals, which therefore do not exhibit nonequilibrium states. Magnetic coupling of nonequilirium subsystems in semiconductors or organic conductors could also result in very slow relaxation because this coupling is relatively weak and macroscopic moments are involved into the system dynamics. Electromagnetic waves and tunnel currents could be advised to vary relaxation rates in films of semiconductors and conducting polymers with coupling in analogy with nonequilibrium superconductors, for example \cite {Tin,Bul}.      

The general GL-type approach to negative magnetization in magnetic systems with coupling can be universally applied to different magnetic formations like ions or polarons. It was well established from experiments that interatomic circulation of shared covalent electrons in closed elementary contours, for example in the benzene ring $\rm C_6H_6$ of six carbon atoms,  results in expulsion of external magnetic fields  from such multi-atom contours with ring covalent bonds \cite {SK}. Would spontaneous circulation of covalent bonds or shared electrons be arranged in some way in polymer loops with magnetic inclusions, then coupling with localized moments of these inclusions could result in negative remnant moments in the zero-field limit. In principle spontaneous charge density circulation might be expected within high symmetry molecular units (or multi ion clusters) with asymmetrical structures for static bonds. Then such units might tend to gain steady current states with loops of shared electron or with interatomic circulation of dynamical bonds. 

Organic chains may be considered for creation of closed dissipationless loops and line stripes with extremely high conductivity, above $10^{11}S/cm$, in prospective polymers like oxidized atactic polypropylene \cite {Eni}. Negative magnetic moments\cite {Rog} with positive slopes of hysteresis curves and  proximity induced supercurrents \cite {Ionov} were reported for ultra-high conducting polymers. There were no observations so far that negative applied fields in these magnetic polymers reversed  anomalous or strong negative moments  into positive values, as it could be expected for real, weak diamagnetism of organic chains. In our view these ferrimagnetic (in high fields) polymers obey, most probably, the universal rules of coupling with negative spontaneous magnetization and positive, paramagnetic-type responses. In other words, the discovered magnetization reversal in ultraconducting polymers might be associated with coupling of two or more magnetic subsystems with positive susceptibilities. One of these subsystem could be almost ideal fermionic fluid of itinerant electrons (without additional requirements for coherent macroscopic states). Another subsystem might be associated, for example, with localized or slowly mobile (in elastic structures) polaron-type formations of ions and trapped spins. 

One may predict from the proposed phenomenology that the stronger paramagnetic or ferrimagnetic high field responses of ultraconducting polymers with localized moments, the better chances for coupled subsystems to maintain strong magnetization reversal in weak fields or at low temperatures.  
In all cases ultraconducting atactic polypropelene,  polydimethylsiloxane, and polyocthylmethacrylate films revealed new and emerging physics for theory and experiments.  Verifications of  quantized or non-quantized flux in macroscopic superconducting loops with polymer weak links might test electron pairing in question in these systems. 

Another promising organic conductor for investigation of possible magnetic coupling is the $\lambda$-DNA molecule, which also exhibited proximity induced superconductivity\cite {Kazu} at very low temperatures. Benzene-type diamagnetism of aromatic rings in the DNA base pairs can screen at room temperatures slight paramagnetic coupling of conducting and  interbase circulating electrons in the double-stranded helix. Comparative low temperature measurements of double- and single-stranded DNA samples could reveal different magnetization and possible GL biomagnetic memory states in $\lambda$-DNA or other nuclear acids.

\section {Conclusions}

Being focused on introduction of a general phenomenological approach to arbitrary condensed matter with magnetic coupling, we had no particular goals in this note to investigate  any immediate applications. However, one may reiterate that coupled magnetic subsystems at convenient temperatures can exhibit memory states under gauss-scale control fields, rather than under kilogauss or tesla fields in conventional magnetic memory systems without coupling (but with an anisotropy axis). Therefore coupling is worth being studied first of all for low-field memory applications, because it might be significant for new recording methods. In fact, the GL model for magnetically coupled nonequilibrium subsystems predicts very strong nonlinearity of magnetic moments in very weak magnetic fields, while such nonlinearity vanishes in moderate and strong fields.

Experimental studies of homogeneous compounds with magnetic coupling can specify at room temperatures prospective nonlinear materials in which negative magnetization might be controlled just by weak field or current pulses. Investigations of nonstationary responses might be especially useful to study nonequilibrium subsystems which are responsible for anomalous magnetic phenomena. Low field magnetic devices can be useful in microelectronics in order, for example, to decrease recording energy flows, to simplify diagnostic medical tools, etc.

We gratefully acknowledge stimulating communications with Dr. A. Junod (Gen$\rm\grave{e}$ve), Dr. K. Mitsen (Moscow), Dr. K. Shambrook (San Francisco).

\end {document}